\documentclass[manuscript,authorversion]{acmart}

\copyrightyear{2022}
\acmYear{2022}
\setcopyright{acmlicensed}
\acmConference[RecSys '22]{Proceedings of the 16th ACM Conference on Recommender Systems}{September 18--23, 2022}{Seattle, WA, USA}
\acmBooktitle{Proceedings of the 16th ACM Conference on Recommender Systems (RecSys '22), September 18--23, 2022, Seattle, WA, USA}
\acmPrice{15.00}
\acmDOI{XXXXXXX.XXXXXXX}
\acmISBN{978-1-4503-8037-9/21/07}

\acmSubmissionID{1238}

\usepackage{bm}
\usepackage[ruled,linesnumbered,commentsnumbered,vlined]{algorithm2e}
\usepackage{graphicx}
\usepackage{subcaption}

\begin{document}


\title{Forgetting Fast in Recommender Systems}

\author{Wenyan Liu}
\authornote{Wenyan Liu and Juncheng Wan are joint first-authors with equal contribution. The research was done when they were both working as interns at ByteDance AI Lab.}
\email{wyliu@stu.ecnu.edu.cn}
\affiliation{%
  \institution{East China Normal University}
  \city{Shanghai}
  \country{China}
}

\author{Juncheng Wan}
\authornotemark[1]
\email{junchengwan@apex.sjtu.edu.cn}
\affiliation{%
  \institution{Shanghai Jiao Tong University}
  \city{Shanghai}
  \country{China}
}

\author{Xiaoling Wang}
\email{xlwang@cs.ecnu.edu.cn}
\affiliation{%
  \institution{East China Normal University}
  \city{Shanghai}
  \country{China}
}

\author{Weinan Zhang}
\email{wnzhang@sjtu.edu.cn}
\affiliation{%
  \institution{Shanghai Jiao Tong University}
  \city{Shanghai}
  \country{China}
}

\author{Dell Zhang}
\authornote{Dell Zhang is the corresponding author. The research was done when he was on leave from Birkbeck, University of London and working full-time at ByteDance AI Lab.}
\email{dell.z@ieee.org}
\orcid{0000-0002-8774-3725}
\affiliation{%
  \institution{Birkbeck, University of London}
  \city{London}
  \country{UK}
}

\author{Hang Li}
\email{lihang.lh@bytedance.com}
\affiliation{%
  \institution{ByteDance AI Lab}
  \city{Beijing}
  \country{China}
}

\renewcommand{\shortauthors}{Liu and Wan, et al.}



\begin{abstract}
Users of a recommender system may want part of their data being deleted, not only from the data repository but also from the underlying machine learning model, for privacy or utility reasons. 
Such \emph{right-to-be-forgotten} requests could be fulfilled by simply retraining the recommendation model from scratch, but that would be too slow and too expensive in practice. 
In this paper, we investigate fast \emph{machine unlearning} techniques for \emph{recommender systems} that can remove the effect of a small amount of training data from the recommendation model without incurring the full cost of retraining. 
A natural idea to speed this process up is to fine-tune the current recommendation model on the remaining training data instead of starting from a random initialization.
This \emph{warm-start} strategy indeed works for \emph{neural} recommendation models using standard \emph{1st-order} neural network optimizers (like AdamW).
However, we have found that even greater acceleration could be achieved by employing \emph{2nd-order} (Newton or quasi-Newton) optimization methods instead.
To overcome the prohibitively high computational cost of 2nd-order optimizers, we propose a new recommendation unlearning approach \texttt{AltEraser} which divides the optimization problem of unlearning into many small tractable sub-problems.  
Extensive experiments on three real-world recommendation datasets show promising results of \texttt{AltEraser} in terms of consistency (forgetting thoroughness), accuracy (recommendation effectiveness), and efficiency (unlearning speed).
To our knowledge, this work represents the first attempt at fast \emph{approximate} machine unlearning for state-of-the-art neural recommendation models.
\end{abstract}

\maketitle

\section{Introduction}

Modern recommender systems rely on \emph{machine learning} techniques such as deep neural networks to model the complex interactions between users and items (e.g., movies, products, news stories), and thus are able to predict what items a user may like.  
The learning of users' personal preferences is made possible by collecting, analyzing, and internalizing their past behavior data.
However, sometimes, it may be necessary or desirable for a recommendation model to intentionally \emph{forget} some of its training data.

There are basically two kinds of motivations for inducing such a controlled ``amnesia'' in recommender systems: \emph{privacy} and \emph{utility}~\cite{chenRecommendationUnlearning2022}.
First, recent studies have found that it is possible for a recommendation model to leak out users' sensitive information being employed in its training~\cite{zhangMembershipInferenceAttacks2021}.
Hence, users may want to remove their sensitive data from the trained model and thus completely avoid such risks to their privacy.
Second, the utility of a recommendation model could deteriorate rapidly due to the noise getting into its training.
To name a few, the training data may contain out-of-date instances, out-of-distribution instances, and polluted instances from poisoning attacks.
Hence, users may want to remove the impacts of such dirty data or bad data from the trained model, so as to improve their experience with the recommender systems.
For example, a user who was recently dumped by his girlfriend would probably prefer not to see recommendations for Valentine's day gifts. 
For another example, a user's son borrowed her mobile phone and watched several YouTube gaming videos, then afterward she would probably get recommendations of the latest computer games which are irrelevant (if she is not really into computer gaming herself) unless the data records for that period of history could be eliminated from the trained model upon her request.

From the perspective of recommender system providers, they cannot ignore users' \emph{right-to-be-forgotten} requests backed by the recent legislation (including 
European Union's GDPR\footnote{\url{https://gdpr.eu/} and \url{https://eur-lex.europa.eu/eli/reg/2016/679/oj}}, 
California's CCPA\footnote{\url{https://leginfo.legislature.ca.gov/faces/billTextClient.xhtml?bill_id=201720180AB375}}, and 
Canada's PIPEDA\footnote{\url{https://www.priv.gc.ca/en/privacy-topics/privacy-laws-in-canada/the-personal-information-protection-and-electronic-documents-act-pipeda/}}),
otherwise they may face significant compliance penalties.

Of course, we can always delete the specific to-be-forgotten data first and then just \emph{retrain} the recommendation model completely from scratch on the remaining data using the same machine learning algorithm as before.
However, such a na\"ive approach to forgetting is obviously a very expensive and time-consuming process that is impractical for real-life large-scale recommender systems.
Are we able to remove the effect of a small amount of training data from the recommendation model quickly without incurring the full cost of retraining?
This is the problem of \emph{machine unlearning}~\cite{bourtouleMachineUnlearning2021}, aka \emph{data removal/deletion}~\cite{guoCertifiedDataRemoval2020} or \emph{selective forgetting}~\cite{golatkarEternalSunshineSpotless2020}, which is receiving more and more attention from the machine learning research community as well as the AI industry.

\begin{figure*}[htb]
    \centering
    \begin{subfigure}[b]{0.45\textwidth}
        \includegraphics[width=\textwidth]{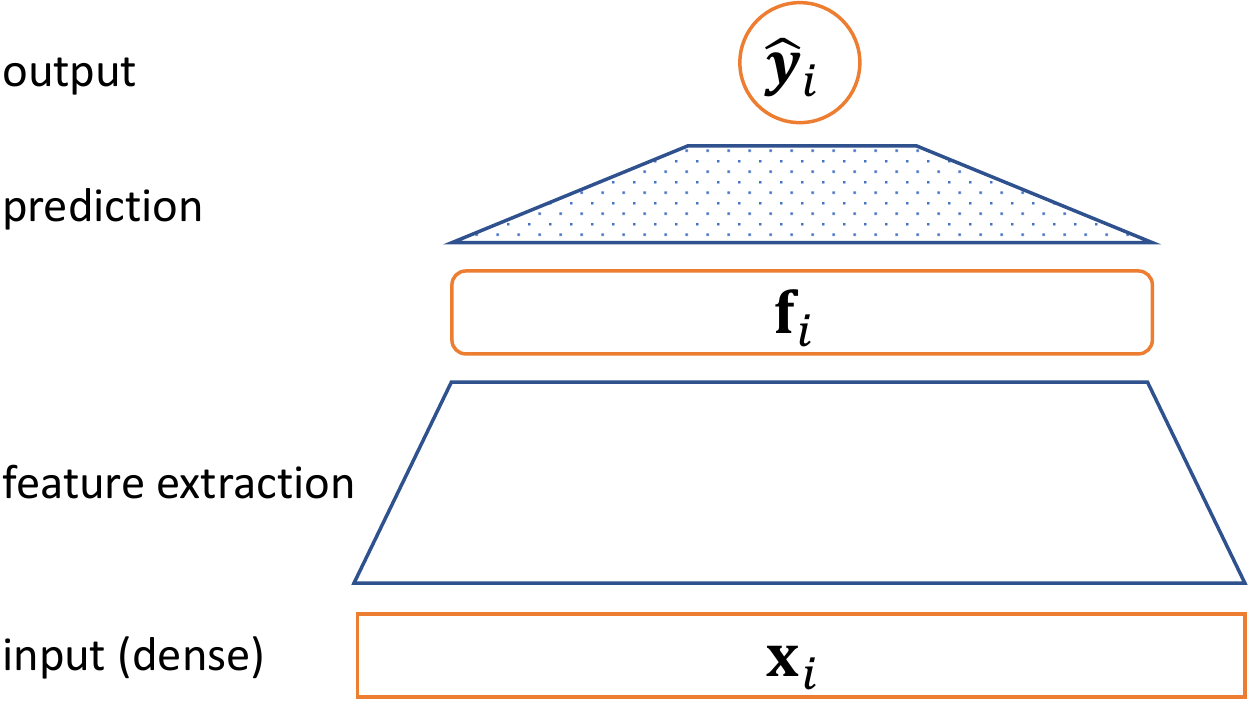}
        \caption{classification}
        \label{fig:neural_classification}
    \end{subfigure}
    \hfill
    \begin{subfigure}[b]{0.45\textwidth}
        \includegraphics[width=\textwidth]{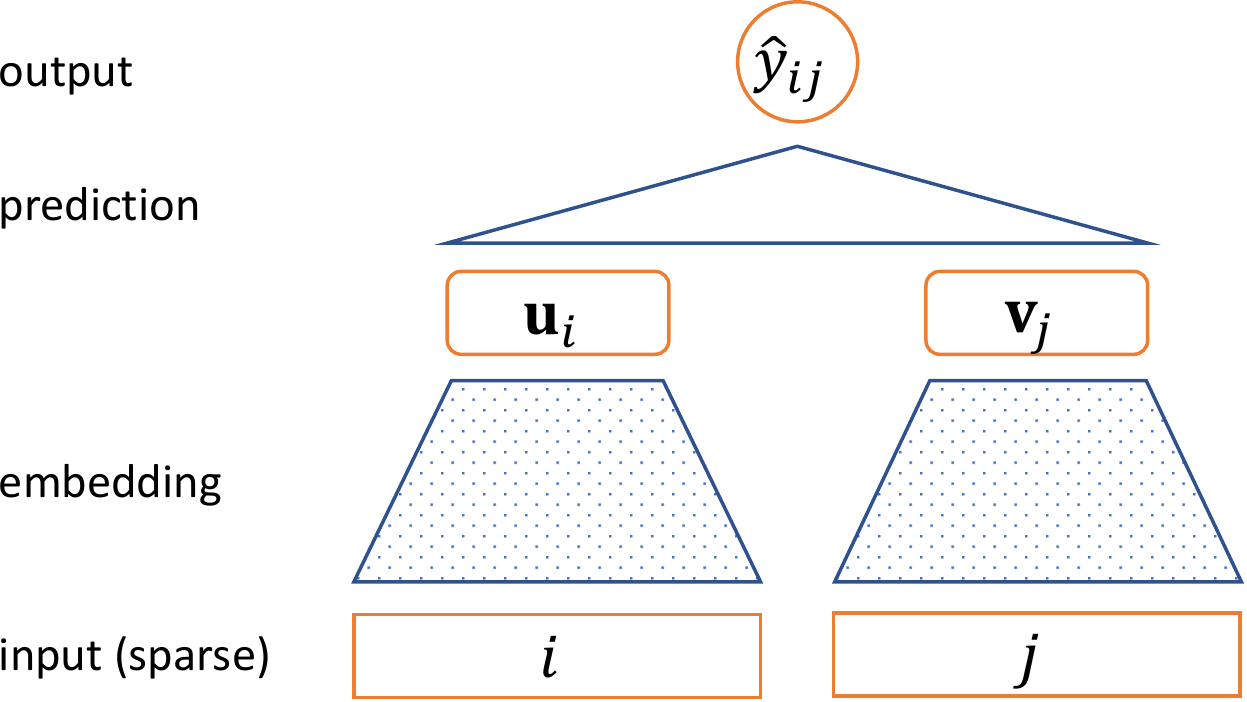}
        \caption{recommendation}
        \label{fig:neural_recommendation}
    \end{subfigure}
    \caption{A comparison of neural network architectures.}
    \label{fig:architecture}
    \Description{A comparison of neural network architectures for classification and recommendation.}
\end{figure*}

Although some investigations into machine unlearning have emerged in the last couple of years, they are mostly devoted to the unlearning of classification (or regression) models in the domains of computer vision and natural language processing (see \S\ref{sec:Machine-Unlearning}).
As illustrated in Figure~\ref{fig:architecture}, there exist fundamental differences in the neural network architectures for classification and recommendation tasks, so directly applying those unlearning methods designed for the former would not work well for the latter, on which we will elaborate later (see \S\ref{sec:Alternating-Optimization}). 
Roughly speaking, to accelerate the unlearning of a neural classification model, one could ``scrub'' only the relatively small prediction layer at the top of the network, while to accelerate the unlearning of a neural recommendation model, we would have to ``scrub'' the huge embedding layers at the bottom of the network.

In fact, the only existing work dedicated to the unlearning of recommendation models (\emph{recommendation unlearning} in short) that we are aware of is \texttt{RecEraser} which appeared a few months ago~\cite{chenRecommendationUnlearning2022}. 
\texttt{RecEraser} belongs to the category of \emph{exact unlearning} methods that hold perfect privacy guarantees but will struggle to cope with forgetting requests in batches, as we will explain in detail later (see \S\ref{sec:Machine-Unlearning}).
It is fairly common in recommender systems that a user would like to remove from the recommendation model, not just one single training instance but a batch of training instances covering a specific period of history. 
To that end, \emph{approximate unlearning methods} which do not insist on absolute privacy guarantees but prioritize the unlearning efficiency would be more suitable.
In this paper, we put more emphasis on utility than on privacy, and thus focus on the area of approximate recommendation unlearning which has not been explored before, as shown in Figure~\ref{fig:taxonomy}.

\begin{figure}[htb]
    \centering
    {\renewcommand{\arraystretch}{2}%
        \begin{tabular}{lp{4cm}p{4cm}}
            \cline{2-3}
            \textbf{Exact \hspace{3mm} Unlearning} &  \multicolumn{1}{|l|}{\texttt{SISA}~\cite{bourtouleMachineUnlearning2021}, \texttt{GraphEraser}~\cite{chenGraphUnlearning2021}, \texttt{DeepObliviate}~\cite{heDeepObliviatePowerfulCharm2021}, etc.} & \multicolumn{1}{l|}{\texttt{RecEraser}~\cite{chenRecommendationUnlearning2022}} \\
            \cline{2-3}
            \textbf{Approx. Unlearning} &  \multicolumn{1}{|l|}{\texttt{Fisher}~\cite{golatkarEternalSunshineSpotless2020}, \texttt{Influence}~\cite{guoCertifiedDataRemoval2020}, \texttt{DeltaGrad}~\cite{wuDeltaGradRapidRetraining2020}, etc.} & \multicolumn{1}{l|}{\texttt{AltEraser} (this paper)} \\
            \cline{2-3}
             & \textbf{Classification/Regression} & \textbf{Recommendation} \\
        \end{tabular}
    }
    \caption{A four-quadrant taxonomy of machine unlearning methods.}
    \label{fig:taxonomy}
    \Description{A four-quadrant of machine unlearning methods.}
\end{figure}

The approach that we propose in this paper, \texttt{AltEraser}, is a new approximate machine unlearning technique tailored to neural recommendation models. 
It further develops the \emph{warm-start} strategy and unleashes the potential of \emph{2nd-order} (Newton or quasi-Newton) optimization methods via an \emph{alternating optimization} framework (see \S\ref{sec:Approach}).
The outstanding performances of \texttt{AltEraser} for approximate recommendation unlearning in three aspects --- consistency (forgetting thoroughness), accuracy (recommendation effectiveness), and efficiency (unlearning speed) --- have been demonstrated by our experiments on several real-world datasets (see \S\ref{sec:Experiments}).

\section{Related Work}
\label{sec:Related-Work}

\subsection{Recommender Systems}
\label{sec:Recommender-Systems}

Collaborative Filtering (CF) is regarded as one of the most popular and successful recommendation techniques because of its simplicity.
Matrix Factorization (MF)~\cite{korenMatrixFactorizationTechniques2009} is among the best CF algorithms driven by the reasonable hypothesis that users with similar past behavior tend to be interested in similar items.
The basic idea of MF is to embed users and items into low-dimensional dense vector space and factorize the sparse user-item interaction matrix into the product of the user and item latent matrices.
MF simulates the observed patterns and completes the missing entries.

Alternating Least Squares (ALS)~\cite{zhouLargeScaleParallelCollaborative2008} is a specific instance of \emph{alternating optimization}~\cite{bezdekConvergenceAlternatingOptimization2003} and can be applied to large maximum-margin matrix factorization (MMMF) problems where it can be reduced to least squares.
ALS itself is a specific algorithm whose target model is fixed which inspires us, while \texttt{AltEraser} is yet another application of the same alternating optimization idea to the general recommendation unlearning problem, with a different motivation.
None of the existing work attempts to approximate unlearning in deep recommendation models.
We cannot use approximate 2nd-order optimization methods directly in recommendation unlearning since adjusting the upper-level parameters is infeasible, and the lower-level embedding accounts for most of the parameters.
The alternating optimization framework helps us split the neural recommendation unlearning into independent, small optimization problems to solve with appropriate initialization and 2nd-order optimization methods.




A representative state-of-the-art recommendation model is Efficient Neural Matrix Factorization (ENMF)\footnote{\url{https://github.com/chenchongthu/ENMF}}~\cite{chenEfficientNeuralMatrix2020}, whose performance is shown to be at least as good as the latest recommendation models, including   
LightGCN~(SIGIR'20), NBPO~(SIGIR'20), LCFN~(ICML'20), DHCF~(KDD'20), and SRNS~(NeurIPS'20).
ENMF resorts to non-sampling learning to solve the inefficiency issue of recommendation models.
Since \texttt{AltEraser} is a general approach agnostic to recommendation models, we choose ENMF as the base model for our experiments.

\subsection{Machine Unlearning} 
\label{sec:Machine-Unlearning}

There are two classes of machine unlearning methods: 
one is \emph{exact unlearning} which completely excludes the to-be-forgotten data in the unlearning process to ensure the perfect privacy protection, and 
the other is \emph{approximate unlearning} which relaxes this stringent requirement so as to attain higher performance.

\subsubsection{Exact Unlearning}
\label{sec:Exact-Unlearning}

There are a few \emph{model-specific} exact machine unlearning methods where the underlying classification model is subject to \emph{decremental updating} (aka \emph{downdating}), e.g., linear Support Vector Machines (SVM)~\cite{tsaiIncrementalDecrementalTraining2014}.
In the context of recommender systems, this type of exact machine unlearning only exists for simplistic recommendation models that could be squeezed into \emph{summation forms}~\cite{caoMakingSystemsForget2015,schelterAmnesiaSelectionMachine2020}, e.g., those based on item popularity.
However, the complexity of most real-world recommender systems is far beyond the reach of such toy models. 



\texttt{SISA} (that stands for Sharded, Isolated, Sliced, and Aggregated) is the seminal work in \emph{model-agnostic} exact machine unlearning~\cite{bourtouleMachineUnlearning2021} which splits the data into disjoint shards and thus reduces the number of model parameters affected by unlearning requests.
\texttt{GraphEraser}~\cite{chenGraphUnlearning2021} is essentially the same unlearning method adapted for graph data with balanced graph partition and a learning-based aggregation method.
\texttt{DeepObliviate}~\cite{heDeepObliviatePowerfulCharm2021} is a similar method based on not data partitioning but model stitching that stores progressive intermediate models during the training process.
When a forgetting request is received, it retrains the overall model in a sequential manner based on those intermediate models. 

\texttt{RecEraser}~\cite{chenRecommendationUnlearning2022} is the existing work closest to this paper, as it translates the data partitioning idea of \texttt{SISA} and \texttt{GraphEraser} of exact machine unlearning from classification to recommendation. 
Our work clearly differs from theirs: while they have focused on exact unlearning to achieve absolute privacy guarantee, we have focused on approximate unlearning to optimize user experience (utility), as shown in Figure~\ref{fig:taxonomy}. 
Although \texttt{RecEraser} is able to significantly outperform \texttt{SISA} and \texttt{GraphEraser}, there is still a noticeable gap between its recommendation accuracy and that of full retraining, when just a single training data point needs to be forgotten.
In that case (forgetting a single data point), our proposed approach \texttt{AltEraser} would have almost the same recommendation accuracy as retraining. 
Therefore, in our experiments, we have tried to remove a lot more training data points from the recommendation model (see \S\ref{sec:Experiments}).
More importantly, the efficiency advantage of such data-partitioning-based exact unlearning methods over retraining would quickly fall apart once the recommender system needs to forget not sequentially arriving individual training data points but a \emph{batch} of training data points at once (which can occur frequently in recommender systems).
Assuming that the training data have been divided into 10 equally-sized partitions (as in the \texttt{RecEraser} paper), each of which resides on a shard, and 30 randomly selected training data instances need to be unlearned.
Thus for any particular shard, the probability that it does not contain any of those to-be-forgotten instances is merely $(1-1/10)^{30} \approx 4\%$, i.e., with about $96\%$ probability its corresponding sub-model will have to be updated. 
This implies that forgetting a reasonably-sized batch of training data instances is likely to force exact unlearning methods back off to training all the 10 sub-models, which would not be very different from complete retraining over the full dataset.
In contrast, the efficiency of our proposed approach \texttt{AltEraser} would be hardly affected by the number of training data points to be forgotten.
As pointed out in some recent papers (including the \texttt{RecEraser} paper), how to efficiently conduct exact unlearning in the batch setting is still an open problem~\cite{bourtouleMachineUnlearning2021,chenRecommendationUnlearning2022}.



\subsubsection{Approximate Unlearning}
\label{sec:Approximate-Unlearning}

The approximate unlearning algorithms developed by Golatkar et al.~\cite{golatkarEternalSunshineSpotless2020} and Guo et al.~\cite{guoCertifiedDataRemoval2020} are most related to this paper, as they also try to utilize 2nd-order optimization techniques to speed up the unlearning process of classification/regression models. 
However, their algorithms cannot be applied to the problem of recommendation unlearning directly due to the sheer scale of the parameter space and the high non-convexity of the recommendation model (see \S\ref{sec:Alternating-Optimization}). 

\section{Problem}
\label{sec:Problem}

\subsection{Neural Matrix Factorization}
\label{subsec:NMF}

Consider a general recommendation problem in collaborative filtering.
Specifically, we try to find low-rank factors of a given user-item interaction matrix $\mathbf{Y} \in \mathbb{R}^{m \times n}$ ($m$ and $n$ are the numbers of users and items, respectively) where some of the values are missing (unobserved). 
This process is related to \emph{matrix completion}. 
One simple method for matrix completion is to use the user matrix $\mathbf{P} \in \mathbb{R}^{m \times d}$ and item matrix $\mathbf{Q} \in \mathbb{R}^{n \times d}$ to model user-item interaction by inner products in a joint latent space (with dimension $d$). 
Each user-item entry $y_{uv}$ is estimated as $\hat{y}_{uv} = \mathbf{p}_u^{\top} \mathbf{q}_v$, where $\mathbf{p}_u$ and $\mathbf{q}_v$ are the $i$-th row of $\mathbf{P}$ and the $j$-th row of $\mathbf{Q}$ respectively. It is proved useful to assign varying confidence levels for each user-item summation term $(y_{uv} - \mathbf{p}_u^{\top} \mathbf{q}_v)^2$. 
This is done by weighted regression with a fixed hyperparameter matrix $\mathbf{W}\in \mathbb{R}^{m \times n}$.
Hence, the objective of this method is to minimize the following loss function
\begin{eqnarray}
    \mathcal{L}(\mathbf{P}, \mathbf{Q}) 
    & = & 
    \frac{1}2\left\|\mathbf{W} \odot (\mathbf{Y}-\mathbf{P}\mathbf{Q}^{\top})\right\|_{F}^2 \nonumber \\
    & = & 
    \frac{1}{2}\sum_{u=1}^{m}\sum_{v=1}^{n} w_{uv}(y_{uv} - \mathbf{p}_u^{\top} \mathbf{q}_v)^2
\end{eqnarray}
where $\|\cdot\|_{F}$ denotes the Frobenius norm. Notations used in this paper are listed in Table \ref{tab:Notations}.

\begin{table}[htb]\small
    \caption{Key Notations Used in This Paper}
    \label{tab:Notations}
      \begin{tabular}{lp{7cm}}
          \toprule
          \textbf{Symbol} & \textbf{Description} \\
          \midrule
          $U$ & The set of $m$ users \\
          $V$ & The set of $n$ items \\
          $\mathbf{Y} \in \mathbb{R}^{m \times n}$ & The user-item interaction matrix with $m$ users and $n$ items \\
          $\mathbf{W} \in \mathbb{R}^{m \times n}$ & The weight matrix indicating the confidence in each interaction \\
          $\mathbf{P} \in \mathbb{R}^{m \times d}$ & The user embeddings for $m$ users \\
          $\mathbf{Q} \in \mathbb{R}^{n \times d}$ & The item embeddings for $n$ items \\
          $\mathbf{p}_u \in \mathbb{R}^d$ & The $u$-th user embedding ($1 \le u \le m$) \\
          $\mathbf{q}_v \in \mathbb{R}^d$ & The $v$-th item embedding ($1 \le v \le n$) \\
          $\mathbf{x}_i \in \mathbb{R}^{2d}$ & The user \& item embedding vector corresponding to the $i$-th interaction ($1 \le i \le N$) \\
          $\bm{\Phi}$ & The model parameters other than $\mathbf{P}, \mathbf{Q}$  \\
          $\bm{\Theta}$ & The neural recommendation model consisting of $\mathbf{P}, \mathbf{Q}$ and $\bm{\Phi}$ \\
          $\mathcal{L}_{D}(\bm{\Theta})$ & The loss of the model $\bm{\Theta}$ on the dataset $D$ \\ 
          \bottomrule
      \end{tabular}
  \end{table}



In general, $w_{uv}=1$ are for existing entries.
As for assigning weights for missing data, there are several common strategies. One straightforward way is to assign a uniform weight $w_0 \in [0, 1]$ on all missing entries.
Nevertheless, in a real recommender system scenario, it is believed that an active user (or a popular item) is more likely to be negative to other items (or users) if there is no interaction. 
From this heuristics, non-uniform weighting is proposed, such as determining $w_{uv}$ with user activity or item popularity measured by frequency.

In the basic Matrix Factorization (MF) model described above, the estimation score $\hat{y}_{uv}$ is linear dependent on embeddings $\mathbf{p}_u$ and $\mathbf{q}_v$. 
Neural Collaborative Filtering (NCF)~ \cite{heNeuralCollaborativeFiltering2017} is modeled with non-linearities by replacing vector products with neural networks. For instance, for one-layered neural networks, $\hat{y}_{uv} = \mathbf{h}^{\top}(\mathbf{p}_u \odot \mathbf{q}_v)$, where $\mathbf{h}\in \mathbb{R}^d$ is the prediction layer. 
%
ENMF~\cite{chenEfficientNeuralMatrix2020} further improves NCF using efficient optimization without sampling. 
In this work, we adopt this MF architecture as the backbone model.

\subsection{Approximate Recommendation Unlearning}

Suppose that a recommendation model $\bm{\Theta}$ has already been learned from the training dataset $D_\text{a}$ consisting of all the available user-item interactions, i.e., 
\begin{equation}
    \bm{\Theta}_\text{a} = \operatorname*{arg\,min}_{\bm{\Theta}} \mathcal{L}_{D_\text{a}}(\bm{\Theta}) \ .
\end{equation}
The task of \emph{recommendation unlearning} is to make the trained model ``forget'' some training examples $D_\text{f} \subset D_\text{a}$ as if they were never used for training the model. 
Without loss of generality, assume that the first $k$ training data instances in the original dataset need to be deleted from the model $\bm{\Theta}_\text{a}$.

The na\"ive approach to machine unlearning is to simply retrain the model from scratch on the remaining dataset $D_\text{r} = D_\text{a} \setminus D_\text{f}$ until convergence to get a new model $\bm{\Theta}_\text{r}$, i.e.,
\begin{equation}
\bm{\Theta}_\text{r} = \operatorname*{arg\,min}_{\bm{\Theta}} \mathcal{L}_{D_\text{r}}(\bm{\Theta}) \ .
\end{equation}
However, such a full retraining approach would be too slow and too expensive in practice. 
When we use the term ``machine unlearning'', we usually mean the algorithms that can unlearn without incurring the cost of retraining from scratch.
Full retraining provides the ground-truth for machine unlearning: \emph{exact} machine unlearning techniques try to replicate the same effect of retraining perfectly, while \emph{approximate} machine unlearning techniques try to achieve the similar effect of retraining. 
In this paper, we focus on the latter, and let $\bm{\Theta}_\text{u}$ denote the new model obtained by machine unlearning.
Our objective is to find a model $\bm{\Theta}_\text{u}$ close enough to $\bm{\Theta}_\text{r}$ as fast as possible.

\section{Approach}
\label{sec:Approach}

\subsection{Warm-Start}
\label{sec:Warm-Start}

Although both full retraining and machine unlearning are essentially optimization problems with the same objective function $\mathcal{L}_{D_\text{r}}(\bm{\Theta})$ to minimize, they have fundamental differences: the former can utilize only the remaining dataset $D_\text{r}$, while the latter has access to the original model $\bm{\Theta}_\text{a}$ and the forgetting dataset $D_\text{f}$ as well. 
There are two basic assumptions about such additional information that we shall rely on to seek more efficient optimization for machine unlearning than full retraining:
(1) $\bm{\Theta}_\text{a} = \operatorname*{arg\,min}_{\bm{\Theta}} \mathcal{L}_{D_\text{a}}(\bm{\Theta})$; 
(2) $|D_\text{f}| \ll |D_\text{a}|$.

Since in machine unlearning, only a small number of training data points need to be forgotten, the optimal unlearned model $\bm{\Theta}_\text{u}$ should be very similar to the original model $\bm{\Theta}_\text{a}$ learned from all training data. 
It is thus a natural idea to employ the \emph{warm-start} (aka \emph{fine-tuning}) strategy to accelerate the training of $\bm{\Theta}_\text{u}$ on the remaining data, i.e., $\bm{\Theta}_\text{u}$ could be initialized not randomly but with $\bm{\Theta}_\text{a}$.

The simplest algorithm for the minimization of a differentiable function is Gradient Descent (GD), aka Steepest Descent.
Its central idea is that $\nabla_{\bm{\Theta}} \mathcal{L}_D(\hat{\bm{\Theta}})$, the gradient of function $\mathcal{L}_D(\bm{\Theta})$,
always points towards the direction of maximal increase at the current point $\hat{\bm{\Theta}}$.
Therefore if we keep moving in the opposite direction of the gradient, we will finally reach a local minimum of the function.

Today, the mainstream optimization algorithm for training a complicated differentiable model like a deep neural network is Stochastic Gradient Descent (SGD), a stochastic approximation of GD which replaces the actual gradient on the entire dataset $D$ by an estimate calculated from a random subset of $D$ (called mini-batch) to reduce the computational burden and accelerate the iterations. 
As mentioned above, the ideal machine unlearning algorithm should be able to quickly update the existing model to $\bm{\Theta}_\text{r} = \operatorname*{arg\,min}_{\bm{\Theta}} \mathcal{L}_{D_\text{r}}(\bm{\Theta})$.
However, SGD would not work well for this purpose. 
As most of the SGD mini-batches for minimizing the updated loss function $\mathcal{L}_{D_\text{r}}(\bm{\Theta})$ would be the same as those for minimizing the original loss function $\mathcal{L}_{D_\text{a}}(\bm{\Theta})$, they do not directly help the unlearning: it would take many, many iterations for the effect of $D_\text{f}$'s removal to eventually show up.

\subsection{Second-Order Optimization}
\label{sec:Second-Order-Optimiation}

The effectiveness and efficiency of the above warm-start strategy could be heavily dependent on the optimization technique that it employs. 
It is well known in the machine learning community that when an initial solution is close to the optimum, 2nd-order optimization techniques such as Newton and quasi-Newton methods are likely to be advantageous due to their fast final convergence~\cite{tsaiIncrementalDecrementalTraining2014}.
Compared with the standard 1st-order optimization methods like SGD or Adam, 2nd-order optimization methods usually take a much longer time in the beginning to go through the first few iterations and find a reasonably good solution, but once they reach the vicinity of the optimum, they tend to converge very quickly (in quadratic or super-linear speed).
This phenomenon has been confirmed by the previous studies as well as our experiments with machine unlearning for classification and regression systems.
The representative approximate machine unlearning algorithms proposed by Golatkar et al.~\cite{golatkarEternalSunshineSpotless2020} and Guo et al.~\cite{guoCertifiedDataRemoval2020} are essentially 2nd-order optimization algorithms making use of the Hessian matrix (or its probabilistic counterpart the Fisher information matrix).
They have been demonstrated to work really well on simple classification or regression datasets~\cite{mahadevanCertifiableMachineUnlearning2021}.

Unlike 1st-order methods which just assume the function surface of $\mathcal{L}_{D_\text{r}}(\bm{\Theta})$ flat, 2nd-order methods try to exploit the function surface's curvature to find a more promising direction for faster optimization of the function. 

 
In essence, the basic 2nd-order optimization method --- \emph{Newton's method} --- for minimizing a function $\mathcal{L}(\bm{\Theta})$ seeks at each iteration an update $\Delta\bm{\Theta}$ which minimizes the 2nd-order Taylor-series approximation of $\mathcal{L}(\bm{\Theta})$ around the current guess $\hat{\bm{\Theta}}$:
\begin{equation}
\mathcal{L}(\hat{\bm{\Theta}} + \Delta\bm{\Theta}) \approx \mathcal{L}(\hat{\bm{\Theta}}) + \nabla_{\bm{\Theta}} \mathcal{L}(\hat{\bm{\Theta}})^\top \, \Delta\bm{\Theta} + \frac{1}2 \, \Delta\bm{\Theta}^\top \, \nabla_{\bm{\Theta}}^2 \mathcal{L}(\hat{\bm{\Theta}}) \, \Delta\bm{\Theta} \ .
\end{equation}
Specifically, with the gradient $\mathbf{g} = \nabla_{\bm{\Theta}} \mathcal{L}(\hat{\bm{\Theta}})$ and the Hessian matrix $\mathbf{H} = \nabla_{\bm{\Theta}}^2 \mathcal{L}(\hat{\bm{\Theta}})$, the minimum of this quadratic approximation is found using the update $\Delta\bm{\Theta} = - \mathbf{H}^{-1} \mathbf{g}$.
This can be shown by taking the derivative of the above quadratic approximation with respect to $\bm{\Theta}$ and setting it to zero.
When the loss function to minimize is indeed quadratic (as in ridge regression), Newton's method is able to reach the global minimum in just one step.
For more complex loss functions, taking a single step of Newton's method can often return a good approximation to the global minimum, especially when the current guess $\hat{\bm{\Theta}}$ is fairly close to the optimal position.

Staring from the original model $\bm{\Theta}_\text{a}$ that should not be far away from the optimal $\bm{\Theta}_\text{u}$, Newton's method for machine unlearning will update it to 
\begin{equation}
\bm{\Theta}_\text{u}^\text{Newton} 
 =  \bm{\Theta}_\text{a} - \mathbf{H}_\text{r}(\bm{\Theta}_\text{a})^{-1} \mathbf{g}_\text{r}(\bm{\Theta}_\text{a}) \ ,
\end{equation}  
where $\mathbf{H}_\text{r}(\bm{\Theta}_\text{a})^{-1} = \nabla_{\bm{\Theta}}^2 \mathcal{L}_{D_\text{r}}(\bm{\Theta}_\text{a})$.
The challenge, of course, is that the above equation for $\bm{\Theta}_\text{u}^\text{Newton}$ is expensive to compute on the large dataset $D_\text{r}$ (whose size is close to the original full dataset $D_\text{a}$), under the assumption that the forgetting dataset $D_\text{f}$ is fairly small. 
The central problem here is how to reduce the computation cost of $\bm{\Theta}_\text{u}^\text{Newton}$.

The above 2nd-order optimization method for machine unlearning could be interpreted as removing the ``influence'' of $D_\text{f}$ from the model $\bm{\Theta}_\text{a}$.
First, because $D_\text{r}$ has only a very small difference from $D_\text{a}$, we could consider using $\mathbf{H}_\text{a}(\bm{\Theta}_\text{a}) = \nabla_{\bm{\Theta}}^2 \mathcal{L}_{D_\text{a}}(\bm{\Theta}_\text{a})$ as a rough replacement for $\mathbf{H}_\text{r}(\bm{\Theta}_\text{a}) = \nabla_{\bm{\Theta}}^2 \mathcal{L}_{D_\text{r}}(\bm{\Theta}_\text{a})$. 
A potential advantage of using $\mathbf{H}_\text{a}(\bm{\Theta}_\text{a})$ instead of $\mathbf{H}_\text{r}(\bm{\Theta}_\text{a})$ is that the former can be computed offline beforehand, though it requires the model to be small enough to allow the explicit storage of the Hessian matrix $\mathbf{H} \in \mathbb{R}^{s \times s}$.  
Second, note that under Assumption (1), $\bm{\Theta}_\text{a}$ is the minimum point of the function $\mathcal{L}_{D_\text{a}}(\bm{\Theta})$, thus in theory the gradient of that function at this point should be zero:
$\nabla_{\bm{\Theta}} \mathcal{L}_{D_\text{a}}(\bm{\Theta}_\text{a}) = \frac{k}{N} \nabla_{\bm{\Theta}} \mathcal{L}_{D_\text{f}}(\bm{\Theta}_\text{a}) + \frac{N-k}{N} \nabla_{\bm{\Theta}} \mathcal{L}_{D_\text{r}}(\bm{\Theta}_\text{a}) = 0$.
This implies $\mathbf{g}_\text{r}(\bm{\Theta}_\text{a}) = \nabla_{\bm{\Theta}} \mathcal{L}_{D_\text{r}}(\bm{\Theta}_\text{a}) = - \frac{k}{N-k} \nabla_{\bm{\Theta}} \mathcal{L}_{D_\text{f}}(\bm{\Theta}_\text{a}) = - \frac{k}{N-k} \mathbf{g}_\text{f}(\bm{\Theta}_\text{a})$.
In other words, the gradient vector $\mathbf{g}_\text{r}(\bm{\Theta}_\text{a})$ over the remaining data $D_\text{r}$ is just the gradient vector $\mathbf{g}_\text{f}(\bm{\Theta}_\text{a})$ over the forgetting data $D_\text{f}$ at the opposite direction and scaled down by a constant factor $\frac{k}{N-k}$. 
Thus, the update of Newton's method at $\bm{\Theta}_\text{a}$ is approximated as 
\begin{eqnarray}
\mathbf{H}_\text{r}(\bm{\Theta}_\text{a})^{-1} \mathbf{g}_\text{r}(\bm{\Theta}_\text{a}) 
& \approx & \mathbf{H}_\text{a}(\bm{\Theta}_\text{a})^{-1} \left[- \left(\frac{k}{N-k}\right) \mathbf{g}_\text{f}(\bm{\Theta}_\text{a})\right] \nonumber \\
& \propto & \mathbf{H}_\text{a}(\bm{\Theta}_\text{a})^{-1}  \, \left[- \sum_{i=1}^k \nabla_{\bm{\Theta}} \ell(\mathbf{x}_i,y_i;\bm{\Theta}_\text{a})\right] \nonumber \\
& = & \sum_{i=1}^k \left[ - \mathbf{H}_\text{a}(\bm{\Theta}_\text{a})^{-1}  \, \nabla_{\bm{\Theta}} \ell(\mathbf{x}_i,y_i;\bm{\Theta}_\text{a}) \right] \nonumber \\
& = & \sum_{(\mathbf{x}_i,y_i) \in D_\text{f}} \mathcal{I}(\mathbf{x}_i,y_i;\bm{\Theta}_\text{a}) \ ,
\end{eqnarray}
where the function $\mathcal{I}(\mathbf{x},y;\hat{\bm{\Theta}}) = - \mathbf{H}_\text{a}(\hat{\bm{\Theta}})^{-1} \, \nabla_{\bm{\Theta}} \ell(\mathbf{x},y;\hat{\bm{\Theta}})$ is exactly the so-called \emph{influence functions}~\cite{kohUnderstandingBlackboxPredictions2017} of the training data point $(\mathbf{x},y)$ with respect to the given model $\hat{\bm{\Theta}}$.




\subsection{Alternating Optimization}
\label{sec:Alternating-Optimization}

Although theoretically appealing, it is infeasible to apply the above 2nd-order optimization techniques directly to the problem of machine unlearning for recommender systems due to the sheer scale of the parameter space and the high non-convexity of the recommendation model.
Even the latest tricks such as \emph{Hessian-free optimization}~\cite{martensDeepLearningHessianFree2010}, \emph{stochastic Neumann series approximation}~\cite{kohUnderstandingBlackboxPredictions2017}, and \emph{Kronecker-factored Approximate Curvature (KFAC)} cannot make 2nd-order optimization computationally affordable for such a massive number of model parameters for user and item embedding, as revealed by our experiments. 

Our key insight comes from the realization that if we conduct the optimization of recommendation model not directly but via the \emph{alternating optimization}~\cite{bezdekConvergenceAlternatingOptimization2003} framework as in ALS~\cite{zhouLargeScaleParallelCollaborative2008}, the huge optimization problem could be divided into many small sub-problems independent of each other and each of those sub-problems is susceptible of 2nd-order optimization with limited computation. 
This paves way for unleashing the potential of 2nd-order optimization in approximate recommendation unlearning.

The \texttt{AltEraser} approach to approximate recommendation unlearning that we propose here iteratively ``scrubs'' the user embeddings $\mathbf{P}$ and the item embeddings $\mathbf{Q}$, in an alternating fashion, so as to remove the effects of $D_\text{f}$ (the training data that need to be forgotten).
The pseudo-code is depicted in Algorithm \ref{alg:AltEraser}.


\begin{algorithm}[htb]\small
\caption{Alternating Eraser (\texttt{AltEraser}) for Approximate Recommendation Unlearning.}
\label{alg:AltEraser}
\DontPrintSemicolon
\newcommand{\forcond}{$i=0$ \KwTo $n$}
\SetKwFunction{AltEraser}{AltEraser}%
\SetKwFunction{OneErasePass}{OneErasePass}%
\SetStartEndCondition{ }{}{}%
\SetKwProg{Fn}{def}{\string:}{}%
\SetKwFunction{Range}{range}%
\SetKw{KwTo}{in}\SetKwFor{For}{for}{\string:}{}%
\SetKwIF{If}{ElseIf}{Else}{if}{:}{elif}{else:}{}%
\SetKwFor{While}{while}{:}{fintq}%
\renewcommand{\forcond}{$i$ \KwTo\Range{$n$}}
\AlgoDontDisplayBlockMarkers\SetAlgoNoEnd\SetAlgoNoLine%
\KwIn{The original user-item data matrix $D_{\text{a}}$, the set of data to be forgotten $D_{\text{f}} \subset D_{\text{a}}$, the original model $\bm{\Theta}_{\text{a}} = \langle\mathbf{P},\mathbf{Q},\bm{\Phi}\rangle$ that consists of user embeddings $\mathbf{P}$, item embeddings $\mathbf{Q}$, and the other model parameters $\bm{\Phi}$.}
\KwOut{The unlearned model $\bm{\Theta}_{\text{u}}$.}
\BlankLine
\Fn{\AltEraser{$D_{\text{a}}$, $D_{\text{f}}$, $\bm{\Theta}_{\text{a}}$}}{
    $D_\text{r}$ $\leftarrow$ $D_\text{a} \setminus D_\text{f}$ \tcp*[h]{The remaining data}\;
    $\widetilde{\mathbf{P}}$, $\widetilde{\mathbf{Q}}$ $\leftarrow$ $\mathbf{P}$, $\mathbf{Q}$ \tcp*[h]{Initialization with warm-start}\;
    \tcp{One unlearning pass over the forgetting users/items}
    $U_\text{f}$,  $V_\text{f}$ $\leftarrow$ the set of users/items occurred in $D_\text{f}$\;
    $\widetilde{\mathbf{P}}$, $\widetilde{\mathbf{Q}}$ $\leftarrow$ \OneErasePass{$U_\text{f}$, $V_\text{f}$, $D_{\text{r}}$, $\widetilde{\mathbf{P}}$, $\widetilde{\mathbf{Q}}$, $\bm{\Phi}$}\;
    \tcp{One or more unlearning passes over all the remaining users/items}
    \While{stopping criterion is not met}{
        $U_\text{r}$,  $V_\text{r}$ $\leftarrow$ the set of users/items occurred in $D_\text{r}$\;
        $\widetilde{\mathbf{P}}$, $\widetilde{\mathbf{Q}}$ $\leftarrow$ \OneErasePass{$U_\text{r}$, $V_\text{r}$, $D_{\text{r}}$, $\widetilde{\mathbf{P}}$, $\widetilde{\mathbf{Q}}$, $\bm{\Phi}$}\;
    }
    \KwRet{\emph{$\bm{\Theta}_{\text{u}} = \langle\widetilde{\mathbf{P}},\widetilde{\mathbf{Q}},\bm{\Phi}\rangle$}}\;
}
\BlankLine
\Fn{\OneErasePass{$U$, $V$, $D_{\text{r}}$, $\widetilde{\mathbf{P}}$, $\widetilde{\mathbf{Q}}$, $\bm{\Phi}$}}{
    \For{\emph{user $u$ (or a mini-batch of users) in $U$}}{
        Hold $\widetilde{\mathbf{Q}}$ and $\bm{\Phi}$ fixed\;
        Update $\widetilde{\mathbf{p}}_u$ in $\widetilde{\mathbf{P}}$ to minimize $\mathcal{L}_{D_\text{r}}(\widetilde{\mathbf{p}}_u,\widetilde{\mathbf{Q}},\bm{\Phi})$ using a 2nd-order optimizer for unlearning\;
    }
    \For{\emph{item $v$ (or a mini-batch of items) in $V$}}{
        Hold $\widetilde{\mathbf{P}}$ and $\bm{\Phi}$ fixed\;
        Update $\widetilde{\mathbf{q}}_v$ in $\widetilde{\mathbf{Q}}$ to minimize $\mathcal{L}_{D_\text{r}}(\widetilde{\mathbf{P}},\widetilde{\mathbf{q}}_v,\bm{\Phi})$ using a 2nd-order optimizer for unlearning\;
    }
    \KwRet{\emph{$\widetilde{\mathbf{P}}$, $\widetilde{\mathbf{Q}}$}}\;
}
\end{algorithm}

For the 2nd-order optimizer in \texttt{OneErasePass}, we could adopt the Hessian Free (HF) Newton~\cite{martensDeepLearningHessianFree2010} method, which implements a number of computational tricks including Hessian vector product, conjugate gradient, backtracking line search, and Levenberg-Marquardt damping.
Besides, we also implemented the standard Newton method with the Hessian matrix being calculated by the manually derived formula (as in~\cite{mahadevanCertifiableMachineUnlearning2021}), which is named Ad Hoc (AH) Newton.

In Algorithm~\ref{alg:AltEraser}, we have only tried to update the user embeddings $\mathbf{P}$ and the item embeddings $\mathbf{Q}$, but not the other model parameters $\bm{\Phi}$.
Our observation indicates that there exists a sufficient degree of freedom within just $\mathbf{P}$ and $\mathbf{Q}$ for the purpose of recommendation unlearning (see \S\ref{sec:Experiments}).
Therefore, we have skipped adjusting $\bm{\Phi}$ in the algorithm specification and our experiments.
However, this is not really a restriction.
If necessary, we can simply add an extra statement to each erasing pass that updates $\bm{\Phi}$ while holding $\mathbf{P}$ and $\mathbf{Q}$ fixed. 










\section{Experiments}
\label{sec:Experiments}

\subsection{Data and Code}

We have conducted our experiments on three publicly-available real-world datasets for recommendation: 
MovieLens-1m\footnote{\url{https://grouplens.org/datasets/movielens/1m/}}, 
Amazon-14core\footnote{\url{http://deepyeti.ucsd.edu/jianmo/amazon/categoryFilesSmall/Electronics_5.json.gz}}, 
and 
KuaiRec-binary\footnote{\url{https://chongminggao.github.io/KuaiRec/}}~\cite{gaoKuaiRecFullyobservedDataset2022}.
Among them, MovieLens is a well-known classic which has been extensively studied, while KuaiRec is a fairly new one that just emerged earlier this year.
Table~\ref{tab:Datasets} shows the basic statistics of these datasets.

\begin{table}[htb]\small
    \caption{Statistics of the datasets.}
    \label{tab:Datasets}
    \begin{tabular}{l|rrrr}
        \toprule
        \textbf{Dataset} & \textbf{\#user} & \textbf{\#item} & \textbf{\#interaction} & \textbf{density} \\ 
        \midrule
        \textbf{MovieLens-1m}   & 6,040 & 3,706 & 1,000,209 & 4.47\% \\
        \textbf{Amazon-14core}  & 1,435 & 1,522 &    35,931 & 1.65\% \\
        \textbf{KuaiRec-binary} & 1,411 & 3,276 &   217,175 & 4.70\% \\
        \bottomrule
    \end{tabular}
\end{table}


The proposed \texttt{AltEraser} has been implemented using Python~3 and PyTorch on top of the open-source recommendation library RecBole\footnote{\url{https://recbole.io/}}~\cite{zhaoRecBoleUnifiedComprehensive2020}. 
We use ENMF as the base recommendation model with L2 regularization (see \S\ref{subsec:NMF}).
The complete code for reproducing our experiments will be made available online\footnote{\url{https://github.com/to-be-released}} for research purposes.

For the standard 1st-order optimization algorithm that is used to train or retrain recommendation models, we have tried several PyTorch built-in optimizers and settled on 
AdamW~\cite{loshchilovDecoupledWeightDecay2019}.
For the 2nd-order optimization algorithm that is employed by \texttt{AltEraser}, we have tried a number of techniques, including 
BFGS, L-BFGS, Conjugate Gradient (CG), Newton CG (NCG), Trust-Region NCG, Generalized Lanczos Trust-Region Newton (Krylov), Gauss-Newton, and AdaHessian.
In the end, two methods --- Hessian Free (HF) Newton~\cite{martensDeepLearningHessianFree2010} and Ad Hoc (AH) Newton~\cite{mahadevanCertifiableMachineUnlearning2021} --- stood out as the best-performing ones overall (see \S\ref{sec:Alternating-Optimization}). 

\subsection{Metrics and Results}

We evaluate different unlearning methods from three perspectives: consistency (forgetting thoroughness), accuracy (recommendation effectiveness), and efficiency (unlearning speed). 
Each experiment would be repeated 5 times, and each time the dataset would be split randomly with 80\% for training and 20\% for testing \emph{per user}.
Then each time, 64 randomly chosen users would each request to remove some of their training data from the recommendation model.
The experiments use early stopping with the patience of 10 epochs.
We report the experimental results averaged over those 5 runs here. 

Let $n_i^+$ denote the number of (implicit feedback) training examples for the $i$-th user. 
For one scenario where users would like to forget private training examples to protect their privacy, we assume that each forgetting user would randomly select
$\lfloor n_i^+/2 \rceil$ out of their training examples and ask the recommender system to delete them.
For the other scenario where users would like to forget noisy training examples to improve the utility, we simulate the noise by randomly injecting
$\lfloor n_i^+/2 \rceil$ false (positive) interactions to each forgetting user and require the recommender system to delete all such noise.
Due to the space limit, only the experimental results for the latter setting are shown in this paper; the omitted experimental results for the former setting exhibit the same patterns and trends, except for the anticipated difference in accuracy (which will be explained later).

\subsubsection{Consistency}

\textbf{Does unlearning forget the specified training data diligently?}
The bottom line is that unlearning should achieve the similar effect of forgetting as retraining (albeit with a fraction of the computational cost).

One possible way to check whether the unlearned model has really met the request to forget is to let the unlearned model make predictions on those training instances which are supposed to be forgotten, as in the previous studies on certifiable machine unlearning for classifiers~\cite{mahadevanCertifiableMachineUnlearning2021}.
Intuitively, if a machine learning model can still remember the labeled training examples that it has seen before, it will probably be able to predict their labels with substantially higher accuracy than in the situation when it has forgotten them.
In the former situation, the model can simply rely on its memory, while in the latter, it will have to make guesses, and some guesses, though educated, may turn out to be incorrect.
In the context of recommendation, this means that a model retaining the memory of those interactions (in question) is likely to assign high probabilities to all of them, but a model without such memory will produce low probabilities for some of them. 
In Table~\ref{tab:Consistency}, we can see that this is indeed the case.
On those interactions which should be forgotten, the average probability output (named ``re-predict score'') of the unlearned models (obtained by either warm-start or \texttt{AltEraser}) would be very similar to the average probability output of the retrained model which is about 3--5 times lower than that produced by the original model.

In addition, we would like to verify how closely the unlearned model approximates the retrained model which is known to have no recollection of any training instance to be forgotten.
For this purpose, we could compare the top-$k$ recommendation list generated by the unlearned model with that produced by the retrained mode using Rank-Biased Overlap (RBO)~\cite{webberSimilarityMeasureIndefinite2010}.
RBO is a relatively new \emph{rank correlation coefficient} designed for information retrieval applications. 
It overcomes several shortcomings of popular rank correlation coefficients such as Kendall's $\tau$ and Spearman's $\rho$.
Specifically, RBO accounts for top-weightedness (by imposing a stronger penalty for differences at the top of the rankings), incompleteness (by handling the top-$k$ lists containing different items without assuming underlying conjointness), and indefiniteness (by limiting the weight of unseen items in the conceptually infinite tail).
RBO values range between 0 and 1, with 1 indicating that the given two top-$k$ ranked lists are identical.
Table~\ref{tab:Consistency} shows the RBO@$k$ (with $k$=10, 20, 50) scores averaged over all the forgetting users.
Here ``Retrain'' represents a recommendation model retrained on the remaining data from scratch and used as the \emph{gold standard} for unlearning, while ``Retrain*'' is yet another recommendation model retrained using the same random initialization but a \emph{different random seed} for the training process.
Although the RBO@$k$ scores for ``Retrain'' itself are always 1, the RBO@$k$ scores for ``Retrain*'' would be substantially lower.
This is because a complex neural network model is likely to have a number of different local minima for its loss function, and thus small variations to its training process would often lead to very different model parameters and outputs, though their predictive performances are similar to each other.
This phenomenon has been observed and reported by many previous studies in deep learning.
Nevertheless, we can see in Table~\ref{tab:Consistency} that our unlearning methods would exhibit very similar RBO@$k$ scores as ``Retrain*'' (with differences less than $0.05$), but the original model which makes no effort to forget would have substantially lower RBO@$k$ scores.
This suggests that our unlearning methods are indeed emulating retraining and working hard to remove the effects of any training data which should be forgotten.   
The experimental results on the test set demonstrate the same pattern that \texttt{AltEraser} and na\"ive warm-start are very similar to ``Retrain*'' in terms of RBO@$k$ scores.



\begin{table*}[htb]\small
    \caption{Consistency (Forgetting Thoroughness) --- Forgetting Users, Training Set}
    \label{tab:Consistency}
    \begin{tabular}{l|c|ccc}
        \toprule
        \textbf{MovieLens-1m}   & Re-Predict Score & RBO@10 & RBO@20 & RBO@50 \\
        \hline
        Original                & 0.4643 & 0.5567 & 0.5771 & 0.5802 \\
        Retrain                 & 0.1318 & 1.0000 & 1.0000 & 1.0000 \\
        Retrain*                & 0.1315 & 0.6434 & 0.6607 & 0.6648 \\
        Warm-Start              & 0.1266 & 0.6984 & 0.7160 & 0.7189 \\
        AltEraser [HF-Newton]   & 0.1390 & 0.7211 & 0.7368 & 0.7395 \\
        AltEraser [AH-Newton]   & 0.1390 & 0.7217 & 0.7370 & 0.7397 \\
        \midrule
        \textbf{Amazon-14core}  & Re-Predict Score & RBO@10 & RBO@20 & RBO@50 \\
        \hline
        Original                & 0.2840 & 0.7619 & 0.7783 & 0.7742 \\
        Retrain                 & 0.0515 & 1.0000 & 1.0000 & 1.0000 \\
        Retrain*                & 0.0513 & 0.8476 & 0.8601 & 0.8590 \\
        Warm-Start              & 0.0543 & 0.8902 & 0.8986 & 0.8977 \\
        AltEraser [HF-Newton]   & 0.0558 & 0.8865 & 0.8941 & 0.8931 \\
        AltEraser [AH-Newton]   & 0.0557 & 0.8870 & 0.8948 & 0.8938 \\
        \midrule
        \textbf{KuaiRec-binary} & Re-Predict Score & RBO@10 & RBO@20 & RBO@50 \\
        \hline
        Original                & 0.6037 & 0.5406 & 0.5490 & 0.5502 \\
        Retrain                 & 0.2356 & 1.0000 & 1.0000 & 1.0000 \\
        Retrain*                & 0.2353 & 0.6499 & 0.6567 & 0.6587 \\
        Warm-Start              & 0.2362 & 0.6424 & 0.6501 & 0.6518 \\
        AltEraser [HF-Newton]   & 0.2499 & 0.6889 & 0.6991 & 0.7004 \\
        AltEraser [AH-Newton]   & 0.2501 & 0.6886 & 0.6988 & 0.7002 \\
        \bottomrule
    \end{tabular}
\end{table*}

\subsubsection{Accuracy}

\textbf{Can unlearning maintain the recommendation effectiveness?}
Forgetting some training data upon users' request must not accidentally ruin the recommendation model's memory of the remaining data.  

We measure the predictive accuracy of recommendation models on the test set using Recall@$k$ and NDCG@$k$ (with $k$=10, 20, 50), which are both widely used in the evaluation of recommender systems~\cite{chenRecommendationUnlearning2022}.

Table~\ref{tab:Accuracy-Forgetting} shows the experimental results of recommendation effectiveness. 
It is clear that all the unlearning methods (including the simple warm-start and the different variants of \texttt{AltEraser}) achieve the same level of performance as retraining: their Recall@10 or NDCG@10 scores would only differ by a negligible amount (less than $0.01$). 
Moreover, it can be seen that for the forgetting users, the unlearned models would provide obviously more accurate recommendations than the original model.
For example, on MovieLens-1m, unlearning with \texttt{AltEraser} [HF-Newton] would bring 16\% and 22\% performance improvements according to Recall@10 and NDCG@10 respectively.
The recommendation effectiveness for the other users would be no worse according to our experiments, though those results are omitted due to the space limit. 
This further proves that unlearning is indeed able to remove the noise in the training data and thus improve the users' experience with the recommender system.


In the other setting where users request to delete not the noise but their true interaction data, we have obtained very similar experimental results except that the unlearned (or retrained) model would have a little bit lower predictive accuracy than the original model, which is understandable because the amount of useful training data has been reduced. 

\begin{table*}[htb]\small
    \caption{Accuracy (Recommendation Effectiveness) --- Forgetting Users, Test Set}
    \label{tab:Accuracy-Forgetting}
    \begin{tabular}{l|ccc|ccc}
        \toprule
        \textbf{MovieLens-1m}   & Recall@10 & Recall@20 & Recall@50 & NDCG@10 & NDCG@20 & NDCG@50 \\
        \hline
        Original                & 0.1449 & 0.2281 & 0.3939 & 0.2937 & 0.2964 & 0.3418 \\
        Retrain                 & 0.1591 & 0.2587 & 0.4425 & 0.3428 & 0.3455 & 0.3916 \\
        Warm-Start              & 0.1606 & 0.2595 & 0.4428 & 0.3490 & 0.3500 & 0.3945 \\
        AltEraser [HF-Newton]   & 0.1671 & 0.2630 & 0.4374 & 0.3573 & 0.3529 & 0.3926 \\
        AltEraser [AH-Newton]   & 0.1674 & 0.2632 & 0.4375 & 0.3581 & 0.3531 & 0.3929 \\
        \midrule
        \textbf{Amazon-14core}  & Recall@10 & Recall@20 & Recall@50 & NDCG@10 & NDCG@20 & NDCG@50 \\
        \hline
        Original                & 0.0621 & 0.1013 & 0.1765 & 0.0531 & 0.0681 & 0.0924 \\
        Retrain                 & 0.0736 & 0.1149 & 0.1883 & 0.0603 & 0.0763 & 0.1005 \\
        Warm-Start              & 0.0701 & 0.1094 & 0.1912 & 0.0568 & 0.0721 & 0.0987 \\
        AltEraser [HF-Newton]   & 0.0710 & 0.1122 & 0.1866 & 0.0572 & 0.0726 & 0.0967 \\
        AltEraser [AH-Newton]   & 0.0710 & 0.1118 & 0.1866 & 0.0571 & 0.0723 & 0.0966 \\
        \midrule
        \textbf{KuaiRec-binary} & Recall@10 & Recall@20 & Recall@50 & NDCG@10 & NDCG@20 & NDCG@50 \\
        \hline
        Original                & 0.1837 & 0.2344 & 0.2912 & 0.2106 & 0.2257 & 0.2474 \\
        Retrain                 & 0.1972 & 0.2485 & 0.3104 & 0.2461 & 0.2521 & 0.2715 \\
        Warm-Start              & 0.1965 & 0.2486 & 0.3057 & 0.2398 & 0.2477 & 0.2650 \\
        AltEraser [HF-Newton]   & 0.1959 & 0.2478 & 0.3063 & 0.2391 & 0.2489 & 0.2667 \\
        AltEraser [AH-Newton]   & 0.1957 & 0.2478 & 0.3064 & 0.2387 & 0.2489 & 0.2668 \\
        \bottomrule
    \end{tabular}
\end{table*}

\subsubsection{Efficiency}

\textbf{How fast is the speed of unlearning?}
For each dataset, we measure the \emph{running time} of each method on the same machine\footnote{The machine used for MovieLens-1m and Amazon 14-core experiments has an Intel{\textregistered} Core{\texttrademark} 8-core CPU i7-9700K @ 3.60GHz and a single NVIDIA GeForce RTX 2080 Ti GPU with 11GB memory, while the machine used for KuaiRec-binary experiments has an Intel{\textregistered} Xeon{\texttrademark} 64-Core CPU E5-2683 v4 @ 2.10GHz and a single NVIDIA GeForce GTX 1080 Ti GPU with 11GB memory.},
and calculate the relative \emph{speed-up} which is defined as the ratio of retraining time to the unlearning time. 

Table~\ref{tab:Efficiency} shows the experimental results about unlearning efficiency. 
We can see that retraining the recommendation model from scratch would require similar time as training the original model, while the simple warm-start strategy using the same 1st-order optimizer could yield a moderate acceleration.  
\texttt{AltEraser} with 2nd-order optimization methods does have high efficiency: using HF-Newton is clearly faster than warm-start, and using AH-Newton achieves the biggest acceleration (up to \textbf{18x} faster than retraining).
The speed of \texttt{AltEraser} could be further increased greatly via parallelization with little effort (see \S\ref{sec:Conclusions}).

As an \textbf{\emph{ablation study}}, we have also tried using the default 1st-order optimizer AdamW in the \texttt{AltEraser} framework, and observed that it would not enjoy the same speed-up as using HF-Newton or AH-Newton.
This confirms that the superior efficiency of \texttt{AltEraser} comes mainly from the successful utilization of 2nd-order optimization techniques rather than performing the optimization in the alternating way. 


\begin{table}[htb]\small
    \caption{Efficiency (Unlearning Speed) --- All Users, Training Set}
    \label{tab:Efficiency}
    \begin{tabular}{l|rr}
        \toprule
        \textbf{MovieLens-1m}   & Running-time (seconds) & Speed-up \\
        \hline
        Original                & 309.69 & -    \\
        Retrain                 & 322.06 & 1.00 \\
        Warm-Start              & 193.07 & 1.67 \\
        AltEraser [AdamW]       & 204.88 & 1.57 \\
        AltEraser [HF-Newton]   &  93.53 & 3.44 \\
        AltEraser [AH-Newton]   &  17.42 & 18.49 \\
        \midrule
        \textbf{Amazon-14core}  & Running-time (seconds) & Speed-up \\
        \hline
        Original                & 231.37 & -    \\
        Retrain                 & 247.72 & 1.00 \\
        Warm-Start              & 217.22 & 1.14 \\
        AltEraser [AdamW]       & 199.86 & 1.24 \\
        AltEraser [HF-Newton]   & 159.40 & 1.55 \\
        AltEraser [AH-Newton]   & 124.96 & 1.98 \\
        \midrule
        \textbf{KuaiRec-binary} & Running-time (seconds) & Speed-up \\
        \hline
        Original                & 270.29 & -    \\
        Retrain                 & 168.69 & 1.00 \\
        Warm-Start              &  97.01 & 1.74 \\
        AltEraser [AdamW]       & 119.41 & 1.41 \\
        AltEraser [HF-Newton]   &  38.35 & 4.40 \\
        AltEraser [AH-Newton]   &  23.72 & 7.11 \\
        \bottomrule
    \end{tabular}
\end{table}

\section{Conclusions}
\label{sec:Conclusions}

Amid the prevalence of Responsible AI mandates, it is becoming more and more important to equip recommender systems with the fast unlearning capability to honor users' right-to-be-forgotten requests. 
As far as we know, our proposed approach --- \texttt{AltEraser} --- is the first attempt at fast \emph{approximate} machine unlearning for state-of-the-art \emph{neural} recommendation models.

The recipe of \texttt{AltEraser} for high-performance recommendation unlearning consists of three key ingredients: (i)~warm-start, (ii)~alternating optimization, and (iii)~2nd-order optimization. 
None of these techniques is anything new by itself. 
The main \textbf{contribution} of this paper, though, lies in the \emph{novel} combination of them designed to accelerate machine unlearning in recommender systems: 2nd-order optimization is most advantageous with warm-start and is only feasible under the divide-and-conquer framework of alternating optimization.

While our experiments have demonstrated that \texttt{AltEraser} can deliver impressive performances, a potential \textbf{limitation} is that it may not always work as billed for all recommendation models on all recommendation datasets. 
As other approximate machine unlearning techniques, one assumption which is required by \texttt{AltEraser} but not necessarily true all the time is that the ideal unlearned model is in the immediate vicinity of the current model.  
More large-scale investigations about the conditions under which \texttt{AltEraser} would work well are certainly going to be beneficial.
Nevertheless, we believe that \texttt{AltEraser} is a valuable new tool in the growing arsenal of machine unlearning techniques.

The field of recommendation unlearning is still in the nascent stage. 
Specifically, our proposed \texttt{AltEraser} approach to approximate recommendation unlearning could have a number of promising \textbf{extensions}.  
First and foremost, similar to ALS~\cite{zhouLargeScaleParallelCollaborative2008}, \texttt{AltEraser} gives rise to potentially massive parallelization of unlearning algorithms.
As we have mentioned earlier, under the alternating optimization framework, the erasing update to one mini-batch of users (or items) would be independent of that to another mini-batch (see \S\ref{sec:Alternating-Optimization}), therefore this is a so-called \emph{embarrassingly parallel} problem which should require little effort to distribute the workload to a cluster of computers and reduce the running time dramatically.
Another idea for the further acceleration of \texttt{AltEraser} is to identify the users and items which would be most affected by unlearning and only update their embeddings.
Intuitively, the deletion of a few users' training data would have the biggest impact on themselves and those who are most similar to them, while the influence over many other users could be minimal.
In addition, we have so far mainly looked into the problem of machine unlearning in recommender systems from the perspective of user experience (utility).
The privacy guarantee of approximate machine unlearning algorithms could be strengthened by further injecting a suitable amount of Laplacian or Gaussian noise (as in \emph{differential privacy}~\cite{dworkDifferentialPrivacy2006}) into the unlearned model~\cite{golatkarEternalSunshineSpotless2020,guoCertifiedDataRemoval2020,wuDeltaGradRapidRetraining2020,mahadevanCertifiableMachineUnlearning2021}, which is an interesting direction to explore for future work.
The degree of privacy protection could then be verified by checking whether the unlearned recommendation model could withstand \emph{membership inference attacks}~\cite{zhangMembershipInferenceAttacks2021}.   
Moreover, we have only studied machine unlearning for general (collaborative filtering based) recommender systems so far. 
It would be interesting to extend the investigation to more sophisticated context-aware, session-based, and knowledge-based recommender systems~\cite{zhaoRecBoleUnifiedComprehensive2020}.
The idea of \texttt{AltEraser} could also go beyond recommender systems and find applications wherever the data are in the nature of bipartite or multipartite graphs. 

A fast approximate machine unlearning algorithm for recommender systems like \texttt{AltEraser} would facilitate the \emph{counterfactual} approach to the generation or evaluation of recommendation \textbf{explanations}. 
Here the term counterfactual means ``\emph{Had you not interacted with those items, you would not receive this recommendation}.''
It has been found by a recent user study that counterfactual recommendation explanations are better correlated with real human judgments~\cite{yaoCounterfactuallyEvaluatingExplanations2022}.

For the purposes of open access, the author has applied a \emph{CC BY public copyright license} to any author accepted manuscript version arising from this submission.

\begin{acks}
We appreciate the preliminary investigation into this research problem conducted by Mimee Xu (NYU) during her internship at ByteDance.
We would also like to thank Dr. Yuanshun Yao (ByteDance) and Dr. Chong Wang (ByteDance) for helpful discussions. 
\end{acks}

\clearpage

\bibliographystyle{ACM-Reference-Format}
\bibliography{ref_information-retrieval,ref_machine-unlearning,ref_maths,ref_explainability,ref_privacy,ref_temporary}


\end{document}